# One-step Pulsed Laser Deposition of Metal oxynitride/Carbon Composites for Supercapacitor Application

Subrata Ghosh*, Giacomo Pagani, Massimilano Righi, Chengxi Hou, Valeria Russo, Carlo S. Casari*

*Micro and Nanostructured Materials Laboratory - NanoLab, Department of Energy, Politecnico de Milano, via Ponzio 34/3, Italy*

**Abstract:** Advanced material composite of nanocarbons and metal-based materials provides a synergistic effect to obtain excellent electrochemical charge-storage performance and other properties. Herein, 3D porous carbon-metal oxynitride nanocomposites with tunable carbon/metal and oxygen/nitrogen ratio are synthesized uniquely by simultaneous ablation from two different targets by single-step pulsed laser deposition at room temperature. Co-ablation of titanium and vanadium nitride targets together with graphite allowed us to synthesize carbon-metal oxynitride porous nanocomposite and exploit them as a binder-free thin film supercapacitor electrode in aqueous electrolyte. We show that the elemental composition ratio and hence the structural properties can be tuned by selecting target configuration and by manipulating the ablation position. We investigate how this tuning capability impacts their charge-storage performances. We anticipate the utilization of as-synthesized various composites in a single PLD production run as next-generation active materials for flexible energy storage and optoelectronic applications.

Corresponding authors Email: subrata.ghosh@polimi.it or subrata.ghoshk@rediffmail.com (Subrata Ghosh) and carlo.casari@polimi.it (Carlo S. Casari)

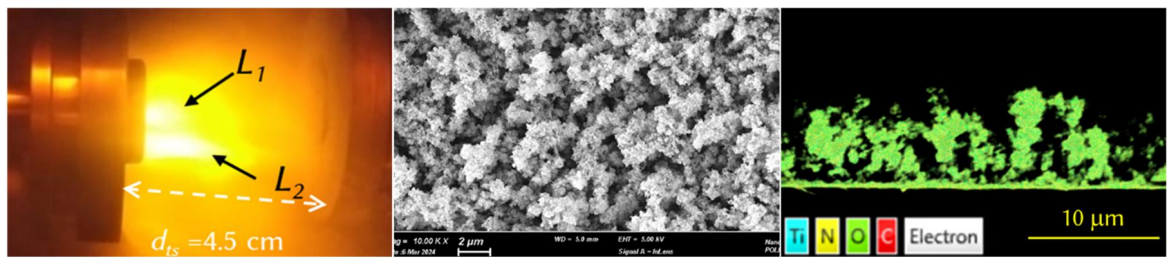

## 1. Introduction:

3D nanostructures consisting of a high surface-to-volume ratio, plenty of active surface sites, and high edge densities, have received significant attention for their potential use in desired applications such as electrochemical energy storage, field emission, and sensors.-[1][2] Considering supercapacitor or electrochemical capacitor technology as an example of a promising application to mitigate the global energy problem, the above-mentioned criteria are to be fulfilled where electrode materials play a pivotal role. Among the electrode materials, carbon-based materials are well-known for excellent power density, good cycle life, and electrochemical stability. They store the charge via double-layer formation at the electrode/electrolyte interface, and the energy storage device is known as electric double-layer capacitor (EDLC). In parallel, metal oxides/nitrides/oxynitrides and conducting polymers offer high specific capacitance but they have poor stability. The corresponding storage device is called pseudocapacitor since fast redox reactions occur at the surface or near the surface of the electrode to store the charge. A great effort has been made to fabricate advanced composite materials combining two types of materials, where one can get a synergistic effect from the individual constituents. For example, a composite of vertical graphene nanosheets decorated with metal oxides,[3] provided around 200 times higher areal capacitance than the vertical graphene alone.[3]

Composite materials can be prepared by one-step and multi-step processes followed by post-treatment.[4][5] In the multi-step process, considering vertical graphene/metal oxide composite as an example,[3] vertical graphene was grown by plasma-enhanced chemical vapour deposition at a relatively high temperature followed by chemical activation to functionalize, metal oxide coating by chemical route such as dip coating and annealing or hydrothermal methods.[2] In this configuration, considering the application in an electrochemical energy storage device, the top surface participates first and then the bottom structure plays a role for the diffusion of electrolytes in the bulk.[2][5] Alternatively, one-step synthesis techniques offer to grow nanostructures with more than one constituent, better bindings between the constituent elements and could offer synergistic effects more effectively while implemented as an active material for desired applications. Taking the electrospinning technique as an example, one can make a homogeneous electrospun solution by adding either metal-based precursors or nanoparticles into the polymer precursor to obtain the composite after electrospinning followed by carbonizations.[6] These multistep processes are time-consuming, needing additional precursors and instrumental facilities, sometimes difficult to control and ensuing non-uniform and non-homogeneous distribution of each constituent.

Now-a-days, the synthesis at room temperature allows to grow nanostructures on a flexible substrate and has received significant attention for its advantage for flexible electronic applications. Although room temperature growth by thin film deposition techniques like sputtering resulted in amorphous materials, these materials also have excellent potential for energy, optoelectronic and other applications.[7][8] The interest in amorphous materials arises due to the availability of a large amount of dangling bonds, defects, free of grain boundaries and electrochemically active surface areas.[8] Among the composite materials, carbon/metal (oxy)nitride has emerged as a promising alternative to the oxide counterpart as it possesses higher electrical conductivity, better stability, and energy storage electrodes.[9] Pulsed laser deposition (PLD) stands out as one of the most versatile techniques, where one can grow varieties of nanostructures with different morphologies starting from compact to porous at room temperature.[10][11][12] However, to prepare a composite material by PLD, sequential ablation of different targets is usually employed leading to one structure on top of another,[10] or a composite target is used to get a composite structure,[11] or simultaneous ablation of multiple targets using multi-beam.[12]

Here, we proposed a unique strategy to prepare a 3D porous carbon/titanium oxynitride and carbon/vanadium oxynitride nanofoam composite by a single-step PLD production run using two different-size targets placing one into another and ablating simultaneously with a single laser beam. We show that we can control the ratio of elemental constituents by flipping the target configuration and by manipulating the laser ablation spot position. As a potential application, the so-obtained 3D nanostructured composites are employed as energy storage electrodes and a symmetric supercapacitor device is assembled, which delivers excellent charge-storage performances.

## 2. Experimental methods
### 2.1. *Synthesis of 3D nanocomposites*

3D nanocomposites were prepared using one-step PLD technique at room temperature. The targets used for this were 1-inch and 2-inches graphite, 1-inch and 2-inches titanium nitride (TiN) and 2-inches vanadium nitride (VN). The laser spot size on the target estimated from the spot obtained on the photosensitive paper at same laser energy and ambient environment is 0.00582 cm$^2$ (Figure S1(a) of supplementary file). For the co-ablation from both targets simultaneously using a single laser, the smaller target was placed on the bigger one and different combinations were carried out to obtain the composite with different constituent elements content (Figure 1a). In this double-target configuration, the ablation spot was captured on the photosensitive paper at the ambient environment to ensure that the laser is falling on both targets (Figure S1b of supplementary file). Silicon, Si (100) were used as substrates for morphological and structural characterization, whereas carbon paper substrates were used to prepare binder-free electrodes for electrochemical tests. The ablation was performed using a Nd:YAG pulsed nanosecond-laser (2$^{nd}$ harmonic at 532 nm, pulse duration 5-7 ns, repetition rate 10 Hz). The deposition was carried out employing two gas pressures:[13][14] 2 Pa for 2 min to grow a compact film as a buffer layer followed by 30 min deposition at 300 Pa to grow the nanofoam composite. The choice of deposition pressure of 300 Pa in the present investigation is due to the fact that deposition at lower pressure (< 200 Pa) led to the compact structure, which may not be suitable for supercapacitor performance. On the other hand, the nanostructure deposited at higher pressure (>300 Pa) is mechanically unstable and showed poor electrochemical charge-storage performance.[15] A similar structure of compact, porous and intermediates can also be obtained by changing the distance between the target and substrate. The distance between the substrate and bottom target ($d_{ts1}$) was 40 mm and the distance between the substrate and top target ($d_{ts2}$) was 34 mm (Figure 1a). All depositions were carried out under nitrogen background gas and the laser pulse energy was about 405 mJ with a fluence of 6.5 J/cm$^2$. Prior to deposition, the chamber was evacuated down to 10$^{-3}$ Pa using rotary pump and turbomolecular pump. The thickness of the nanostructure can be controlled by varying deposition times. However, a micrometer thick film is of our interest for the electrode to use for energy storage applications on a commercial scale. Thus, we have chosen a 30 min deposition time to obtain a micrometer thick nanofoam composite. For the sake of comparison, we prepared pristine TiN nanofoam keeping same process parameters during the deposition.

### 2.2. *Morphological and structural characterization.*

A field-emission scanning electron microscope (FESEM, ZEISS SUPRA 40, Jena, Germany) was employed to investigate the morphology of composite. The volumetric void fraction or porosity of the nanofoam composite was estimated using SEM images by ImageJ software. After setting the scale and cropping the information bar from the SEM micrograph, the image is binarized setting the threshold properly, in such a way that just the pores should be colorful. After this important step, the analysis can be performed to estimate the volumetric void fraction from the option "measure" available in imageJ.[16] Electron Dispersive X-ray spectra (EDX) were recorded at the acceleration voltage of 10 kV using Aztec software to evaluate the elemental analysis and the instrument is equipped with a Peltier-cooled silicon drift detector (Oxford Instruments). The Raman spectra of all samples were recorded by Renishaw Invia Raman spectrometer, UK. The spectra were acquired using 514 nm laser with a power of 0.4 mW, a 1800 gratings spectrometer, a 50× objective lens, and 20 accumulations with accumulation time of 10 s.

**2.3 Electrochemical measurements**.

The electrochemical performances of the nanocomposites were carried out in a 2-electrode configuration using Swagelok Cell (SKU: ANR-B01, Singapore) and 6M KOH used as an electrolyte. The diameter of each electrode is 1 cm. The photographic image of an electrode is provided in Figure S1(c) of supplementary file. The hydrophobic PP membrane was modified by a two-step process: soaked with acetone at room temperature for 5 min and followed by aqueous 6M KOH solution , and used after 1 hr.[17]  The cell was assembled by sandwiching separator-soaked-electrolytes between the nanocomposites grown on carbon paper. Prior to the test, nanocomposite electrodes and a modified separator were dipped into the electrolyte solution for several hours. Cyclic voltammogram, charge-discharge test and electrochemical impedance spectra were recorded using a PALMSENS electrochemical workstation. Before recording the original data, the proto-type cells were scanned at 100 mV/s scan rate for 1000 cycles. The cyclic voltammetry at different scan rates ranging from 20 to 1000 mV/s and charge-discharge at different current densities of 150 to 500 µA were carried out. The areal capacitance was calculated using the equation: $C_{areal} = \int I \, dV / Av\Delta V$ , where *I* is the current, *v* is the scan rate, *A* is the geometric area of the electrode and *ΔV* is the voltage of the device. Single electrode capacitance = 4 × device capacitance. The volumetric capacitance of electrode materials was estimated by dividing the areal capacitance by the total height of two symmetric nanocomposite electrodes. The electrochemical impedance spectroscopy was conducted in the frequency range of 1 Hz to 0.1 MHz at open circuit potential with a 10 mV *a.c.* perturbation. The specific capacitance at 120 Hz is calculated using the relation of $C_s = 1/2\pi f Z'' A$. The relaxation time constant (τ$_0$) was estimated from the corresponding frequency (*f*$_0$) at -45° phase angle using the equation of $f_0 = 1/\tau_0$. The resistor-capacitor time constant at a frequency (*f*) of 120 Hz was calculated using the relation of $\tau_{RC} = Z'/2\pi f Z''$, where *Z'* and *Z"* are the real and imaginary impedance, respectively.

## 3. Result and Discussions:
### 3.1. Carbon-Ti oxynitride composites

The C/Ti oxynitride composite nanofoam (C/TiO$_x$N$_y$) is prepared by placing a 1-inch TiN target on top of a 2-inch graphite target and ablated at 250 Pa of pressure (Figure 1a). The morphology is found to be porous (Figure 1b) with a volumetric void fraction of 81.55% and an average uniform thickness of 6.4 µm (Figure 1c). In our PLD set up, both the target and substrate were rotated during the deposition to obtain the homogeneous and uniform deposition over the area of 2 cm diameter. With the translation of substrate, the homogeneous and uniform deposition on the substrate can be scaled up further. From the cross-sectional image (Figure 1c), the nanofoam composite consists of a compact layer on the substrate with a porous structure on top. The compact layer provides better adhesion with the current collector, excellent conductivity, and complementary capacitance, whereas the porous structure provides a large specific surface area and open porous structures for fast electrolyte ion transport.[18] The basic constituents of C/TiO$_x$N$_y$ from top-view EDX measurement (Figure 1d) are carbon (48.7 at.%), titanium (Ti, 7.7 at.%), nitrogen (N, 16.3 at.%) and oxygen (O, 27.3 at.%). Figure 1(e-h) shows the cross-sectional electron image and corresponding elemental mapping, which confirms the uniform distribution of each element in the nanofoam composite. Although pure nitrogen was used as background gas during the ablation, the presence of inevitable oxygen in the composite is attributed to adventitious oxygen and native thin oxide layer formation on the porous nanocomposite upon air exposure.

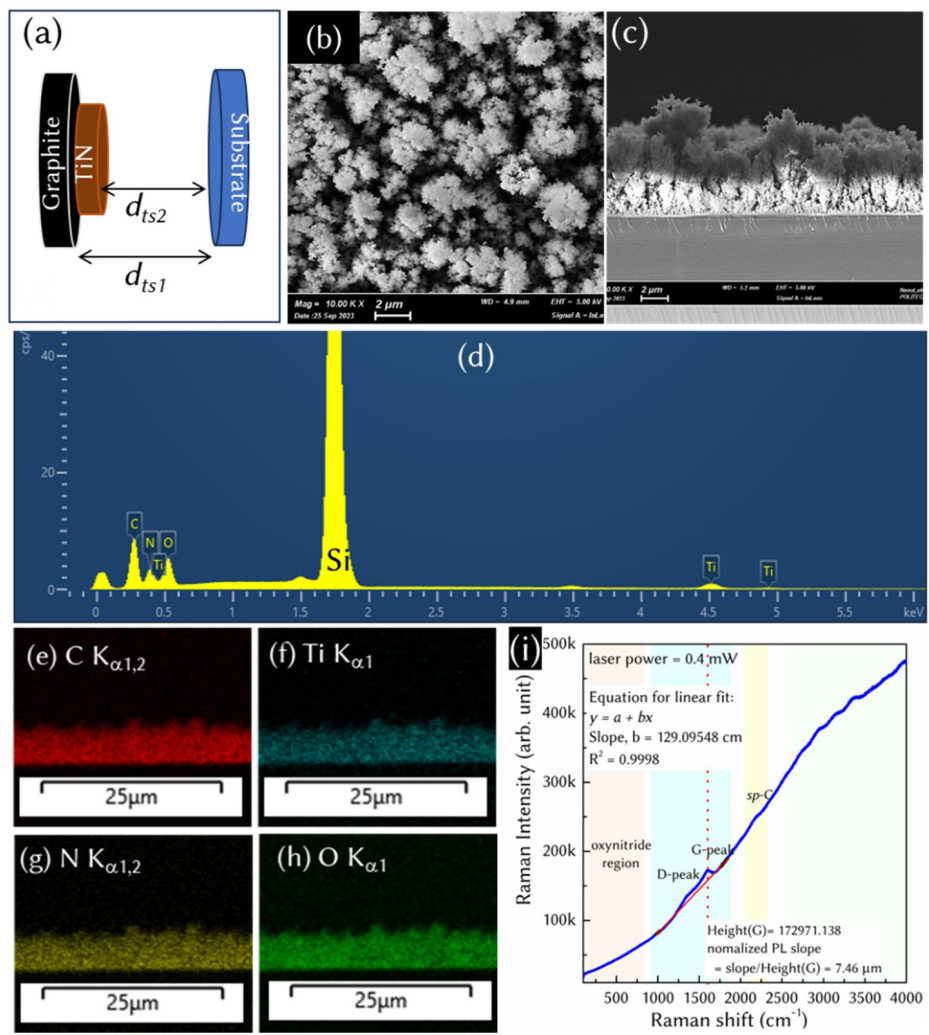

**Figure 1: Morphology and structure of Carbon-titanium oxynitride C-TiO$_x$N$_y$.** *(a) schematic of two-target assembly for the deposition in PLD. (b) Top view and (c) cross-sectional scanning electron micrograph, (d) EDX spectra, (e) electron image, and elemental mapping of (f) carbon, (g vanadium, (h) nitrogen, and (i) oxygen. Raman spectra of nanofoam composite with 514 nm laser*

Figure 1(i) shows the Raman spectrum of C/TiO$_x$N$_y$, which consists of $sp^2$-carbon band with two components namely, D-peak (located at 1372 cm$^{-1}$) and G-peak (located at 1585 cm$^{-1}$). The D-peak is attributed to the breathing mode of hexagonal rings, and it is related to disorder in the structure, and G-peak is due to the carbon-carbon stretching mode of $sp^2$-carbon. Other carbon peaks found were $sp$-carbon located at 2170 cm$^{-1}$ and second orders in the range of 2500 to 4000 cm$^{-1}$. The huge photoluminescence (PL) background, as can be seen from the Raman spectra, is quite dominating and can be attributed to the adsorbed water on the light density and porous carbon-based nanostructures[19] and the presence of metallic Ti-components. In the study of wide range of hydrogenated amorphous carbons, the normalized PL slope is found to be increases exponentially with the hydrogen content.[19] The normalized PL slope is estimated from the ratio of slope in the linear region of 1000-1900 cm$^{-1}$ (in the unit of cm) and height of the G-peak (Figure 1(i)).[19] With the same analogy, to provide a quantitative measure of metallic components' influence in the lightweight porous structure, the normalized PL slope estimated is 7.46 µm, which is much higher than the amorphous carbon nanofoam (3.5-4.3 µm).[20] This huge PL background is also one of the reason behind negligible contribution of oxynitride in Raman spectra.. Since the structure is amorphous, it shows broadened G-peak with full-width a half maximum of 134.6 cm$^{-1}$ (Table 1).

As-grown nanofoam composite on the carbon paper substrate is used as a binder-free electrode in a supercapacitor. The performance of the C/TiO$_x$N$_y$ symmetric device is shown in Figure 2. The cyclic voltammogram (CV) profile for the C/TiO$_x$N$_y$ device is found to be mirror-symmetric and unchanged near-rectangular in shape even at the higher scan rate (Figure 2a). Deviation from perfect rectangular CV can be attributed to the pseudocapacitive contribution from oxynitride and functional groups of carbon along with the double-layer capacitive contribution of carbons. The plot of the areal and volumetric capacitance of C/TiO$_x$N$_y$ with respect to the scan rate is shown in Figure 2b. The areal (volumetric) capacitance of the device is found to be 0.61 mF/cm$^2$ (480 mF/cm$^3$) at the scan rate of 0.1 V/s and it retains 50% at a high scan rate of 1 V/s. The areal capacitance of single electrode of 2.4 mF/cm$^2$ at 0.1 V/s, which is obtained by multiplying the device areal capacitance by 4, is found to be comparable with many reports such as vertical graphene nanosheets (0.47 mF/cm$^2$),[3] Ni$_3$(HHTP)$_2$@woven fabrics (0.205 mF/cm$^2$), TiO$_2$ nanogrid (0.74 mF/cm$^2$)[21] and titanate hydrate nanogrid (0.08 mF/cm$^2$).[21] Moreover, the cycle stability of the device is found to be 86% at 100 µA even after 10000 charge-discharge cycles, indicating good electrochemical stability. The inset of Figure 2c shows the charge-discharge profile of the device for a few cycles. In order to explore the charge-storage kinetics, electrochemical impedance spectroscopic results (Figure 2d and Figure S2 of supplementary file) are investigated through the Nyquist and Bode plot.[22] At 120 Hz, the negative phase angle of the device is found to be 52° and the specific capacitance is estimated to be 91 µF/cm$^2$ with good frequency response (Figure 2d). The obtained specific capacitance of our nanofoam composite at 120 Hz is found to be impressive and comparable withP/N-doped carbon (30 µF/cm$^2$),[23] vertically oriented graphene (87 µF/cm$^2$)[24] B/N-doped carbon (99 µF/cm$^2$) and, laser processed carbon-titanium carbide heterostructure (118 µF/cm$^2$) [18] based devices. On over that, the negative phase angle of our device at 120 Hz is comparable and/or better than N-doped mesoporous carbon (0°)[25], activated carbon-based commercial electric double layer capacitor (0°)[26],[27], thermally reduced graphene oxide (30°)[28], 3D ordered porous graphene film (53°)[29], etc. Moreover, it has been demonstrated for NiTe$_2$-based devices that the phase angle around 53° is sufficient for the AC line filtering purpose.[30] This result suggests that the nanofoam composite not only has good charge-storage properties but also can be used for filtering applications. The elemental composition ratio is manipulated by flipping the targets and then changing the ablation spot to improve the charge-storage performance further, which is detailed in the following.

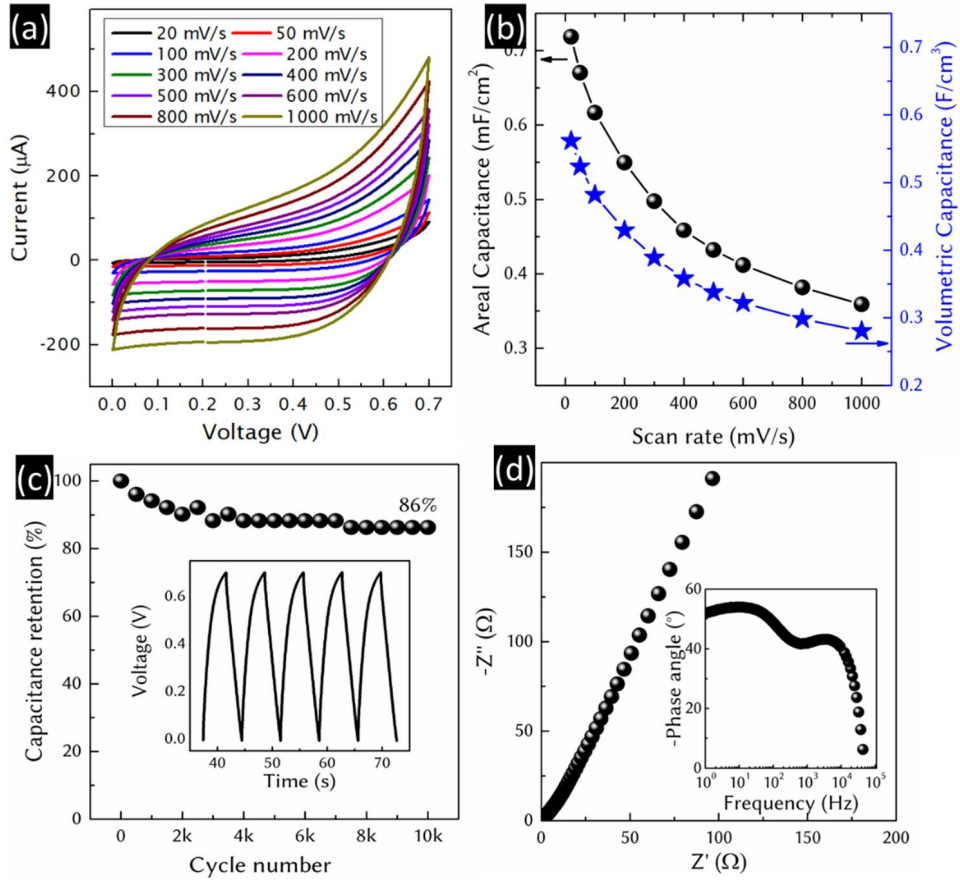

**Figure 2: Supercapacitor performance of Carbon-Titanium oxynitride.** *(a) cyclic voltammogram at different scan rates, (b) areal and volumetric capacitance with respect to the scan rate, (c) cycle stability with the charge-discharge cycles at 100 µA in an inset, (d) Nyquist plot with Bode plot in the inset.*

To explore such possibilities, we flipped the target compared to the previous case by placing the 1-inch graphite target on the 2-inch TiN target (Figure 3a) at first and then changed the position of ablation spot slightly (Figure 3b). The ablation was carried out at 250 Pa pressure keeping other deposition parameters the same. With our set-up, the ablation spot can be changed by 1 mm precisely and depositions are carried out at three different ablation spots. Prior to this, let us discuss the different zones of plasma plume for the case of ablation in a single target (Figure 3c) for better visualization of three deposition processes and their consequences. The plasma plume consists of hot core, mid regions and cold regions followed by shock wavefronts. In the mid-region, where ions and neutral coexist due to the ionization and recombination, we can see two different bright zones for the case of co-ablation: one is coming from the bottom target, and another is from the top (Figure 3d-f). For the sake of simplicity, we define the length of those two bright zones as $L_1$ for the top target and $L_2$ for the bottom one. It can be easily seen that if one changes the ablation spot position slightly, the relative ratio of $L_1$ and $L_2$ can be controlled, which also has an impact on the final structure obtained.

    Case I: the ablation position is more on the carbon side as a result $L_1 > L_2$ (Figure 3d)
    Case II: the ablation position in the middle of the boundary, where $L_1 \approx L_2$ (Figure 3e)
    Case III: the ablation is more on TiN such that $L_1 < L_2$ (Figure 3f)

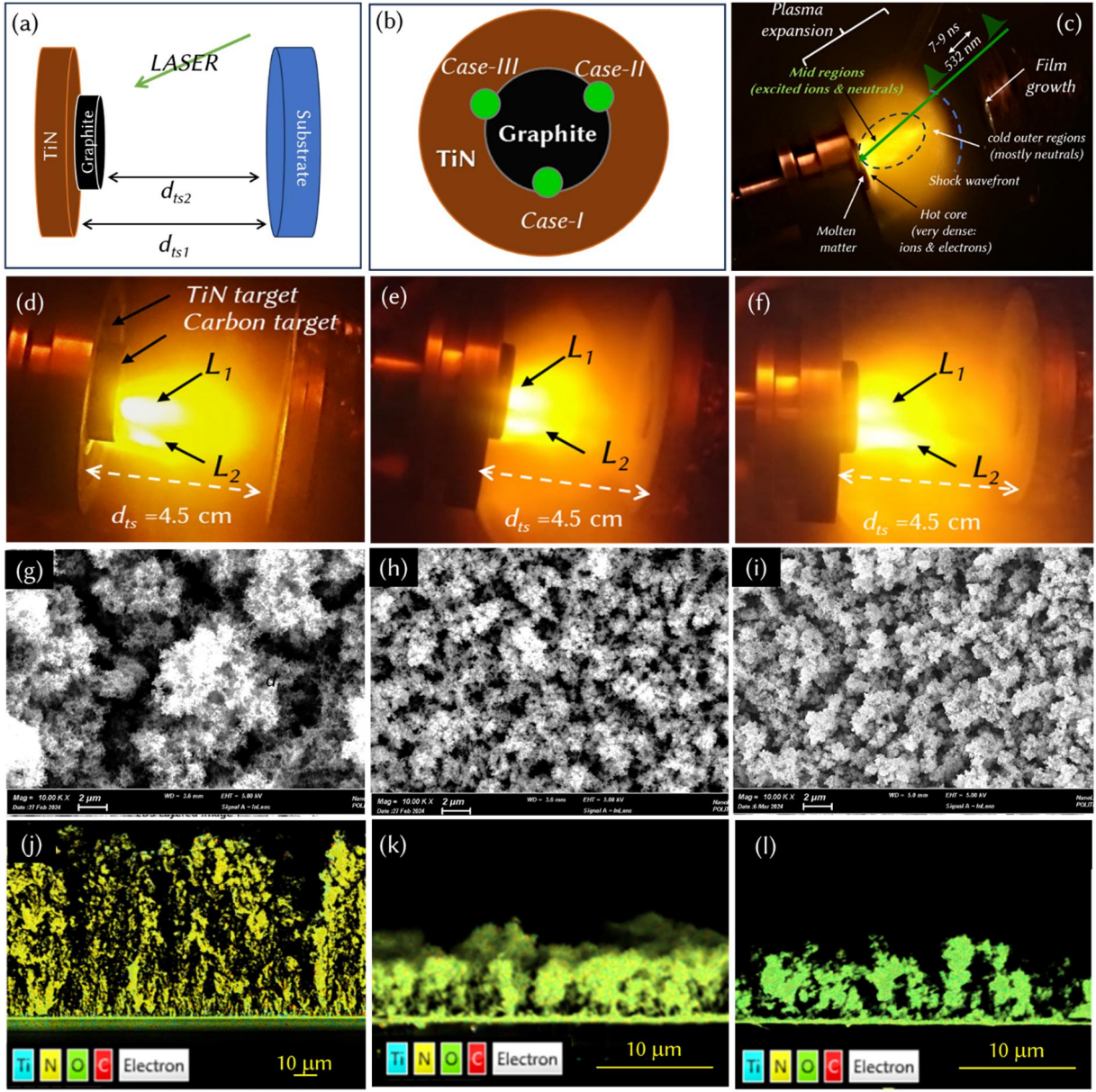

*Figure 3: (a) Schematic of targets assembly and (b) three ablation spots. (c) Photographic image of plasma plume for single target ablation by 532 nm laser demarcating different zones of plasma. (d-f) Representative photographic image of plasma plume for three different ablation conditions. Corresponding (g-i) scanning electron micrographs, (j-l) elemental layered image.*

For $L_1>L_2$ (Figure 3b and 3f), since the ablation spot is more on the graphite target side, the ablation yield of carbon is much higher than that of TiN and $d_{ts\text{-graphite}}$ is less than the $d_{ts\text{-TiN}}$. As a result, the average growth rate of the final composite is found to be 2.34 μm/min (Figure 3j). As the ablation position shifted towards the TiN target, the growth rate was found to be reduced to 0.2 μm/min for the other two cases (Figure 3f-g). The morphology (Figure 3g-i), of $C/TiO_xN_y$-1 is more porous and $C/TiO_xN_y$-3 possesses a relatively higher density of nanoparticles. The volumetric void fraction is found to be decreased from 90% ($C/TiO_xN_y$-1) to 67% ($C/TiO_xN_y$-3) as the ablation spot shifted from the graphite-rich zone to the TiN-rich zone. The maximum thickness of the film for $C/TiO_xN_y$-1, $C/TiO_xN_y$-2, and $C/TiO_xN_y$-3 is found to be 81.2, 8.1 and 7.9 μm, respectively. The corresponding cross-sectional elemental layered image is shown in Figure 3(j-l) and the elemental mapping of each element for all composites is given in Figure S(3-5) in the supplementary file, which confirms the uniform and homogeneous distribution of each elements in the nanofoam. From the elemental compositional analysis, as obtained from the top-view EDS result, C/TiO$_x$N$_y$-1 contains higher carbon content with 86.3 at.% and less Ti-content about 1.4 at.% along with oxygen and nitrogen content, which is obvious as explained before. Whereas, C/TiO$_x$N$_y$-2 and C/TiO$_x$N$_y$-3 contain Ti(carbon) of 22.9 (24.3) and 31.4 (4.0) at.%, respectively since the laser ablation spot was shifted towards the nitride-side maintaining the condition of co-ablation from both targets. Strikingly, the maximum oxygen content (61.1 at.%) has been seen for C/TiO$_x$N$_y$-3 compared to that of C/TiO$_x$N$_y$-2 (43.9 at.%), as a higher amount of Ti is prone to physisorbed atmospheric oxygens more. Although the ablation spot position is same, the higher Ti-content of C/TiO$_x$N$_y$-2 (22.9 at.%) compared to that of C/TiO$_x$N$_y$ (7.7 at.%) is attributed to the arrangement of targets as the TiN target was placed below the graphite target for the former case. For the C/TiO$_x$N$_y$ composite, increasing Ti- and N-content or decreasing carbon content also leads to an increase in the amount of oxygen content. Increasing N-content would improve the electrochemical stability whereas O-content may contribute more pseudocapacitance to the double-layer capacitance when the composite is explored as an energy storage electrode.

The Raman spectra of all composites are shown in Figure 4(a-b). In the Raman spectra of C/TiO$_x$N$_y$-1, the D-peak and G-peak are quite prominent (Figure 4b), and the spectra contain a lower photoluminescence background (normalized PL slope of 1.21 µm) and very weak contribution from TiO$_x$N$_y$ as C/TiO$_x$N$_y$-1 contains high carbon content. In low frequency region (100 to 1000 cm$^{-1}$) of Raman spectra of C/TiO$_x$N$_y$-1 (Figure 4(c)), we observed two large bands, which are compatible with amorphous titanium oxynitride as reported in the literature.[31] Raman spectra of C/TiO$_x$N$_y$-2 (Figure 4a) show D- and G-peak with huge amount of PL background (normalized PL slope of 5.98 µm), which is similar to the C/TiO$_x$N$_y$ (Figure1i) and expected due to the same target assembly except flipping. However, the relative ratio of constituent elements is different for those two cases, which reflects in normalized PL slope, FWHM of G-peak and I$_D$/I$_G$ ratio (table 1). We noticed a different Raman spectra of C/TiO$_x$N$_y$-3 than the other studied nanofoam composite, and hence direct comparison with C/TiO$_x$N$_y$-1 and C/TiO$_x$N$_y$-2 may not be feasible. First of all, the Raman spectrum of C/TiO$_x$N$_y$-3 is dominated by TiO$_x$N$_y$ contribution and less contribution from carbon structure resulted in noisy Raman spectra, particularly in the region of 1000-2000 cm$^{-1}$ (Figure 4a-b), which is in agreement with EDX result. The normalized PL slope is estimated to be 3.66 µm. To know more about it, the Raman spectrum of C/TiO$_x$Ny-3 is recorded for longer accumulations and fitted with four Gaussian lineshapes (the fitted spectrum is provided in Figure S6 of supplementary file) at 1145, 1281, 1489, and 1596 cm$^{-1}$. In most of the metal decorated carbons, the changes associated with metal decoration are the broadening in D- and G-peak and changes in I$_D$/I$_G$ ratio, which is also observed for C/TiO$_x$N$_y$, C/TiO$_x$N$_y$-1 and C/TiO$_x$N$_y$-2. It has been reported for the (Ti, V, Zr, W) doped carbon structure of amorphous carbon films that the changes in Raman intensity with the metal concentration are due to the increased surface reflectivity of the films due to the metal's incorporation in the carbon matrix.[32] In another report of metal-doped amorphous carbon,[33] the additional peak at around 1100 cm$^{-1}$ was observed and attributed to the *sp*$^3$ C-C vibration in amorphous carbon and it was eventually found to be stronger in Fe-doped amorphous carbon than the Ni-doped amorphous carbon. In the report,[33] the D-peak position of Fe-doped a-C:H is also found to be shifted to 1300 cm$^{-1}$ from the pristine a-C:H film (1340 cm$^{-1}$). In the report on Raman spectra of 3% Co-doped carbon thin film,[34] several overlapped Raman peaks have been seen in the range of 1000 to 1800 cm$^{-1}$ and attributed to fullerene-related carbon materials. Moreover, the peaks at 1150 and 1500 cm$^{-1}$ are also ascribed to the trans- and cis-polyacetylene chains in many reports,[35] which is not in the case here as our synthesis is hydrogen-free. A Narrow band is also noticed in carbon onion film between 1400 and 1550 cm$^{-1}$, which is attributed to the heptagon formation.[36] On the other hand, from the study of reduced graphene oxide, those peaks were attributed to the oxygen content and structural disorder. [37][38] Thus, the present Raman spectra of C/TiO$_x$N$_y$-3 certainly indicate the coexistence of different forms of carbons together. In analogy to the metal-doped carbon thin film and the oxygen-adsorption into highly porous C/TiO$_x$N$_y$-3 structure with high content of Ti, the prominent peaks at 1145 and 1489 cm$^{-1}$ are attributed to the TiN-induced modification in amorphous carbon containing short range $sp^2$-C, where the aromatic bond decreases to the detriment of polyacetylene chains, and formation of carbon-onion kind structure.[34] However, further investigation to gain a clear insight could be a subject of research. To know more about the structural changes in the nanofoam composites prepared at different ablation spots, the FWHM of G-peak and I$_D$/I$_G$ are estimated and found to be decreased from the C/TiO$_x$N$_y$-1 to C/TiO$_x$N$_y$-3, which can be attributed to the improvement in the structure of nanofoam composite and different stoichiometry of individual elements (table 1). Thus, it is clear from the Raman spectroscopic analysis that all nanofoam composites very disordered, amorphous and combination of carbon and Ti-oxynitride, and they have different structural quality since they consist of different amount of elemental compositions, which is consistent with the EDX result.

In order to check the reproducibility, we carried out the repeated deposition of C/TiO$_x$N$_y$-3 (C/TiO$_x$N$_y$-3R). The morphology (volumetric void fraction of 65.8%), maximum thickness (~8 μm) and almost similar structural characteristics in terms of Raman spectra is observed for C/TiO$_x$N$_y$-3R (Figure S7 of supplementary file). We also would like to note that continuous erosion of targets for repetitive growths may affect the reproducibility and in that case one needs to tune the ablation spot further.

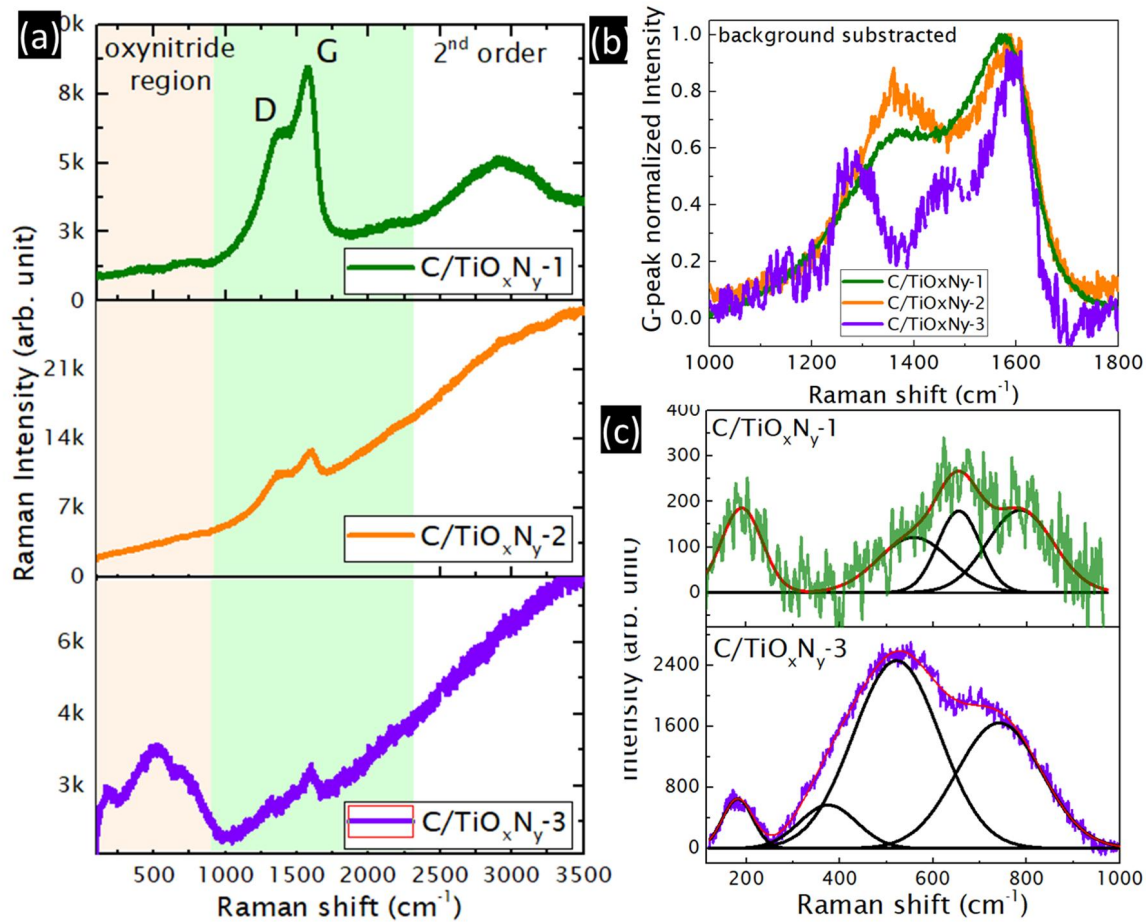

*Figure 4: (a) Visible Raman spectra of C/TiO$_x$N$_y$-1, C/TiO$_x$N$_y$-2, and C/TiO$_x$N$_y$-3. The wavenumber range of 100 to 3500 cm-1 is selected to show the photoluminesce background for all composites. (b) background subtracted and G-peak normalized Raman spectra in the range of 1000 to 1800 cm-1 of the nanofoam composites. To reduce the signal-to-noise ratio, the Raman spectra of C/TiO$_x$N$_y$-3 was recorded for quite long accumulations and smoothened compared to other two. (c) background corrected and fitted Raman spectra of nanofoam composite in the oxynitride zone with deconvoluted peaks.*

The charge-storage performances of all supercapacitor devices using the above three composites were carried out in a symmetric configuration (Figure 5). The cyclic voltammogram at different scan rates and charge-discharge profile at different current densities for all composites are provided in Figure S8 of the supplementary file. A comparative volumetric CV of all three devices at 0.5 V/s is shown in Figure 5(a). Among all, C/TiO$_x$N$_y$-2 supercapacitor device shows the highest area under the CV curve, while normalized with the height of the films (Figure 5(a) and table 1). The highest areal capacitance at 0.1 V/s is obtained for C/TiO$_x$N$_y$-1 (1.85 mF/cm$^2$) supercapacitor with an electrode height of around 75 µm (Figure 5(b)). Whereas the highest volumetric capacitance of 1763 mF/cm$^3$ at 0.1 V/s is obtained from C/TiO$_x$N$_y$-2 device. In order to validate the synergy, the electrochemical performance of bare TiO$_x$N$_y$ nanofoam device is performed. The morphology, Raman spectra and electrochemical performance of bare TiO$_x$N$_y$ nanofoam are provided in Figure S9 of supplementary file. The areal capacitance of bare TiO$_x$N$_y$ nanofoam device at 0.1 V/s is found to be 0.5 mF/cm$^2$, which is lower than that of studied C/TiO$_x$N$_y$ nanofoam composite device (table 1). At given scan rate (0.1 V/s), the volumetric capacitance of C/TiO$_x$N$_y$-2 (1760 mF/cm$^3$) and C/TiO$_x$N$_y$-3 (526 mF/cm$^3$) is found to be higher than that of bare carbon nanofoam symmetric device (215-522 mF/cm$^3$)[20] and bare TiO$_x$N$_y$ nanofoam (480 mF/cm$^3$), which reveals the importance of nanofoam composite. However, the synergy effect on the charge-storage performance of C/TiO$_x$N$_y$ (volumetric capacitance of (480 mF/cm$^3$) and C/TiO$_x$N$_y$-1 device (volumetric capacitance of 124 mF/cm$^3$) (table 1) is not observed, which indicates that the optimized metal oxynitride in carbon structure is needed to obtain the synergy effect in terms of charge-storage performance. Based on the obtained physico-chemical properties of nanofoam composites (table 1), establishing a direct correlation between one physico-chemical parameter and charge-storage performance is not straightforward as all local properties and structures are important for capacitive charge storage.[39] However, it has been seen that the maximum volumetric capacitance is obtained when C/Ti ratio of the sample is close to 1. We anticipate that the synergy effect can be more prominent for the post-treated nanofoam composite with improved structural properties and more insights on the structure-property-performance relationship can be obtained, which is the focus of our future investigation. On the other hand, the C/TiO$_x$N$_y$-3 supercapacitor device shows better rate performance and cycle stability (Figure 5(b-c)). This can be attributed to the composition and its structure as explained in EDX and Raman spectroscopic results.

To probe more insight, the double-layer capacitance and pseudocapacitance of nanofoam composites using Trasatti Method[40][41]. The double layer capacitance is obtained from the intercept of volumetric capacitance versus inverse square root of scan rate plot (Figure S10 of supplementary file). The reciprocal total capacitance is obtained from the intercept of reciprocal volumetric capacitance versus square root of scan rate plot (Figure S10 of supplementary file). Pseudocapacitance contribution is obtained from the difference between total capacitance and double-layer capacitance. The contribution of double-layer capacitance and pseudocapacitance is normalized and plotted in Figure 5c. The charge-storage of nanofoam composite (both C/TiO$_x$N$_y$-1 and C/TiO$_x$N$_y$-2) is mostly dominated by pseudocapacitance (Figure 5c), which may be due to the fact that the nanocomposite contains a huge amount of functional groups (table 1). Amongst, C/TiO$_x$N$_y$-3 showed 35% double-layer capacitive contribution and 65% pseudocapacitive contribution, which could be due to the structural changes as seen from Raman spectra after the electrochemical investigation (Figure 6). In terms of energy storage performance, better rate performance (Figure 5b) and cycle stability (Figure 5d) of C/TiO$_x$N$_y$-3 can be attributed to the balanced double-layer and pseudocapacitive contribution. The impedance spectra of the nanocomposites are shown in Figure 5d and Figure S11 of supplementary file. While looking at the high frequency zone of impedance spectra (Figure 5d), C/TiO$_x$N$_y$-2 supercapacitor device exhibited lower equivalent series resistance, steeper line at low-frequency region, higher phase angle at 120 Hz, smaller relaxation and capacitor-resistor time constant and higher specific capacitance at 120 Hz compared to that of C/TiO$_x$N$_y$-1 and C/TiO$_x$N$_y$-3 superacapacitor (table 1). This result suggests that C/TiO$_x$N$_y$-2 consists of optimized compositional elements and structural quality.

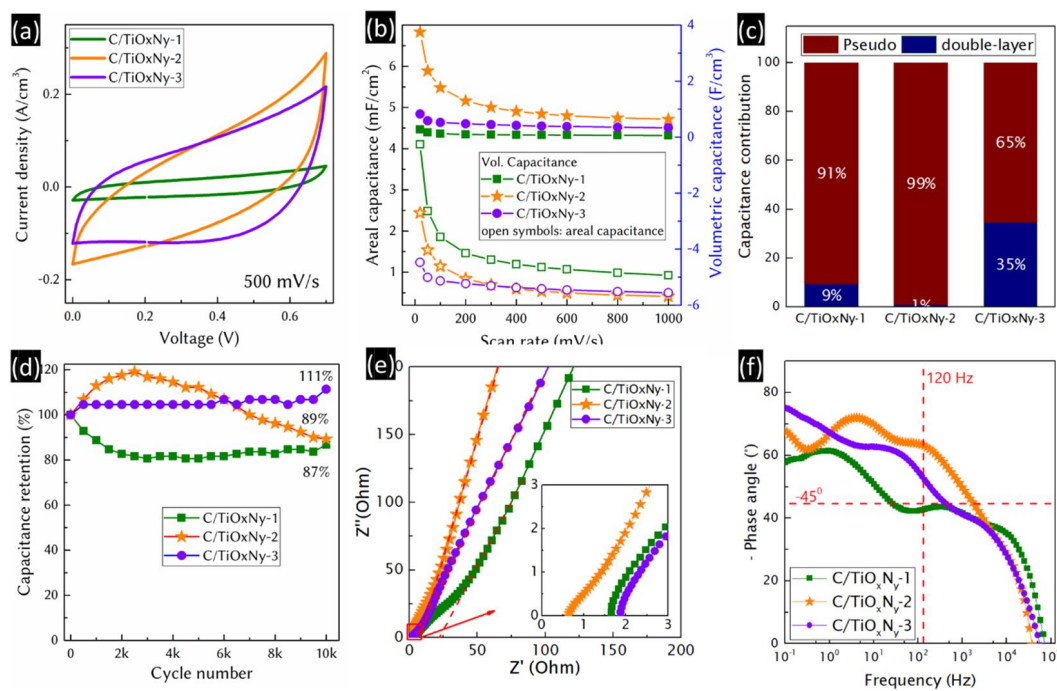

*Figure 5: Supercapacitor performance of Carbon-Titanium oxynitride. (a) volumetric cyclic voltammogram at 0.5 V/s, (b) areal and volumetric capacitance at different scan rates, (c) double-layer and pseudocapacitive contributions estimated using Trasatti Method analysis. (d) cycle stability over 10000 charges-discharges, (e) Nyquist plot with an inset represent Nyquist plot at high-frequency zone demarcating equivalent series resistance obtained from the plot at zero Z'-axis and (f) Bode plot. Dashed lines in each Nyquist plot crossing the zero Z'-axis represent charge-transfer resistance. Dashed red lines in the phase angle represent the characteristic frequency at -45° phase angle.*

As seen from Figure 5d, a rise in capacitance (~120%) was observed in the first 2000 cycles for the C/TiO$_x$N$_y$-2 supercapacitor and more than 100% capacitance retention for C/TiO$_x$N$_y$-3 supercapacitor after 10000 charge-discharge cycles can be attributed to the electrochemical activation and active nitrogen and titanium vacancies. To ensure the electrochemical activation and structural changes in the composite upon prolonged charge/discharge cycles, C/TiO$_x$N$_y$-2 and C/TiO$_x$N$_y$-3 supercapacitors were disassembled, electrodes were washed with water to remove the electrolyte residues and dried in an ambient environment. The Raman spectra of the electrode after electrochemistry are recorded and shown in Figure S12 in supplementary file (C/TiO$_x$N$_y$-2) and Figure 6 (C/TiO$_x$N$_y$-3). A prominent change in spectra after electrochemical investigation is observed for C/TiO$_x$N$_y$-3. Firstly, a huge photoluminescence has been seen in comparison with the Raman spectra of bare nanocomposite due to the electrolyte ion intercalation during the charge-storage process and water adsorption in the structure during the wash. Interestingly, many sharp peaks arise in the oxynitride region of C/TiO$_x$N$_y$-3 after the electrochemical investigation, and the highest position of peaks are 123, 195, 286, 367, 447, 671 and 904 cm$^{-1}$. The peaks obtained from C/TiO$_x$N$_y$-3 after the electrochemical process are similar to the Raman spectra of titanium oxynitride nanofilms,[42] grown at 200 °C and high fluence using 193 nm laser by PLD, deconvoluted into the peaks at 120 cm$^{-1}$ (B$_{1g}$), 281 cm$^{-1}$ (multi-phonon), 412 cm$^{-1}$ (E$_g$), 606 cm$^{-1}$ (A$_{1g}$), 671 cm$^{-1}$ (2$^{nd}$ order), and 766 cm$^{-1}$ (B$_{2g}$) of rutile TiO$_2$, and 355 cm$^{-1}$ (longitudinal acoustic) and 509 cm$^{-1}$ (transverse optical) of TiN. The position of these peaks and TiN contributions in the composite film depends on the deposition parameters. Firstly, since our deposition of nanofoam composites was carried out at room temperature, as-grown structures are amorphous in nature. Moreover, it has been reported that the presence of an acoustic band at around 200-300 cm$^{-1}$ along with an optical band in 500-600 cm$^{-1}$ in amorphous TiO$_x$N$_y$ is associated with the titanium vacancy in the structure,[31] which is also seen in the Raman spectra of C/TiO$_x$N$_y$-3 nanofoam composite (bottom spectra of Figure 4a). On the other hand, in contrast to the Raman spectra of as-grown C/TiO$_x$N$_y$-3 nanofoam, we can see two prominent peaks without additional peaks after the electrochemical investigation. Thus, the transformation of broad Raman bands to several sharp peaks in the oxynitride region after electrochemical investigations is one of the indications of the structural changes in oxynitride due to the electrochemical activation. In addition, the improvement in D- and G-peak along with the evolution of second order region in Raman spectra is noticeable for C/TiO$_x$N$_y$-3 after the electrochemistry, also indicate the electrochemical activation of the structure.

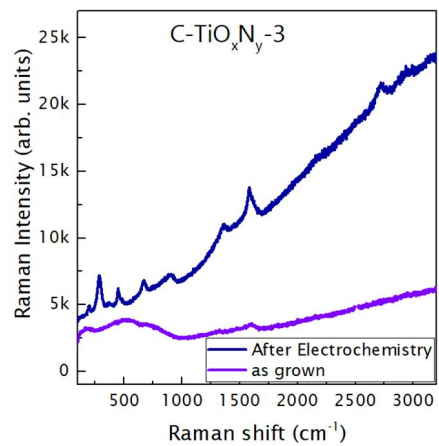

*Figure 6: Raman spectra of C-TiO$_x$N$_y$-3 after electrochemical investigation along with the spectra of as-grown nanofoam.*

*Table 1: Physico-chemical and charge-storage properties of C/MO$_x$N$_y$ composites (M stands for Ti and V). The charge-storage performances were carried out in a 6M KOH electrolyte.*

| Nanofoam composite (C: carbon) | Avg. Height of film (μm), porosity | C/M/N/O (at.% from EDX; M: Ti and V) | Photo-luminescence background (μm) | Full width at half maxima (FWHM) of/G (cm$^{-1}$) | Intensity ratio of D-to-G | Vol. (areal) capacitance in mF/cm$^3$ (mF/cm$^2$), Rate performance, cycle stability after 10k cycles of composite | Area specific Capacitance (μF/cm$^2$) at 120 Hz | Phase angle at 120 Hz | Relaxation time constant in ms (frequency in Hz) at -45° | Capacitor - Resistor time constant (ms) at 120 Hz | Equivalent series resistance (Ω) |
|---|---|---|---|---|---|---|---|---|---|---|---|
| C/TiO$_x$N$_y$ | 6.4, 81.55% | 48.7/7.7/16.3/27.3 | 7.6 | 134.6 | 0.89 | 480 (0.61) at 0.1 V/s, 58.3% at 1 V/s, 86% at 100 μA | 91 | -53° | 4.7 (209.7) | 0.98 | 0.6 |
| C/TiO$_x$N$_y$-1 | 75, 90.02% | 86.3/1.4/3.3/9.0 | 1.21 | 117.0 | 0.82 | 124 (1.85) at 0.1 V/s, 50% at 1V/s, 87% at 125 μA | 78 | -43° | 40 (24.8) | 1.39 | 1.7 |
| C/TiO$_x$N$_y$-2 | 6.5, 88.72% | 24.3/22.9/8.8/43.9 | 5.98 | 111.0 | 0.79 | 1763 (1.14) at 0.1 V/s, 36% at 1 V/s, 89% at 125 μA | 109 | -63° | 0.6 (1643.6) | 0.67 | 0.7 |
| C/TiO$_x$N$_y$-3 | 7.5, 67.13% | 4.0/31.4/3.5/61.1 | 3.66 | 51.9 | 0.77 | 526 (0.78) at 0,1 V/s, 63% at 1 V/s, 111% at 125 μA | 82 | -53° | 2.7 (363.8) | 0.96 | 1.9 |
| C/VO$_x$N$_y$-2 | 18, 60.37% | 46.3/10.7/19.9/23.1 | 2.78 | 151.8 | 0.69 | 260 (0.96) at 0.1 V/s, 48% at 1 V/s, 71% at 100 μA | 132 | -51° | 8.1 (123.3) | 1.3 | 0.6 |

### 3.2. Carbon-V oxynitride composites

As can be seen from previous investigations that C/TiO$_x$N$_y$-2 has better charge-storage performance, we attempted to replace TiN by VN target with same configuration. This approach also helps to showcase the versatility of our strategy. In which, a 1-inch graphite target was placed on VN target with a diameter 2-inch to co-ablate simultaneously at the pressure of 250 Pa and other deposition parameters were the same (Case-II, similar to Figure 3a. See details in experimental section). Moreover, VN nanocrystalline structures has high charge storage capacitance of 1340 F/g at 2 mV/s and it can be operated within the electrochemical stability window of -1.2 to 0 V vs reference electrode in aqueous electrolyte medium.[43] Thus, VN-based became a popular choice to be used as negative electrode in asymmetric supercapacitor device, which is the major boost to obtain higher energy density of supercapacitor device. However, major problem with vanadium nitride nanostructures is electrochemical stability and hence there is urgent need of designing composite with carbon.[9] The 3D porous morphology of nanofoam composite (C/VO$_x$N$_y$-2) is confirmed from the electron micrographs (Figure 7a) with a porosity of 60% and the average height of the film in the uniform region is found to be. Compared to the C/TiO$_x$N$_y$-2 (average height of 6.5 µm), the higher average height of the film of 18 µm (inset of Figure 7a) with the same target-assembly configuration (case-II) is attributed to the higher ablation yields of VN than that of TiN.

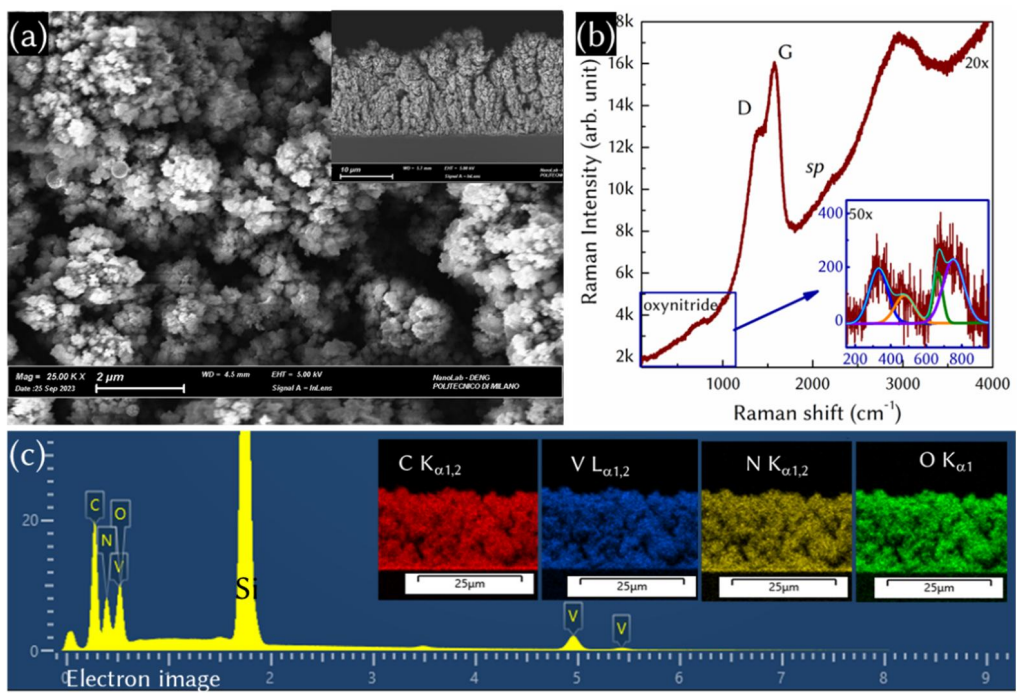

*Figure 7: **Morphology and structure of Carbon-vanadium oxynitride.** (a) Scanning electron micrograph, (b) Raman spectra of nanofoam composite with low-frequency range spectra accumulated for higher time at inset. (c) EDX spectra, (d) electron image, and elemental mapping of (e) carbon, (f) vanadium, (g) nitrogen, and (h) oxygen.*

The Raman spectrum of the film is shown in Figure 7b. It consists of a weak peak in the low-frequency zone of 100 to 1000 cm$^{-1}$, and $sp^2$-carbon peaks with second-order counterpart, and a weak $sp$-carbon peak. The low frequency region consists of two broad bands, which is compatible with amorphous vanadium oxynitride.[44] Thus, the Raman spectra of C/VO$_x$N$_y$ confirms the co-existence of functionalized amorphous nanocarbons and vanadium oxynitride. Figure 7c shows the EDX spectra of C/VO$_x$N$_y$-2, which consists of 46.3 at.% carbon, 10.7 at.% vanadium, 19.9 at.% nitrogen and 23.1 at.% oxygen. The EDX result again confirms that the sputtering yield for C/VO$_x$N$_y$-2 (elements at.% ratio of C/V = 4.33 and V/N = 0.54) is completely different than the C/TiO$_x$N$_y$-2 (elements at.% ratio of C/Ti = 1.06 and Ti/N = 2.6). The corresponding cross-sectional elemental mapping (Figure 7d-g) of C/VO$_x$N$_y$-2 confirms the uniform distributions of each element in the nanofoam composite.

The electrochemical charge-storage performance of the C/VO$_x$N$_y$-2 composite is carried out by fabricating a symmetric cell. The CV of the nanofoam composite is shown in Figure 8a. The maximum areal (volumetric) capacitance of the nanofoam composite device obtained at 0.1 V/s is 0.96 mF/cm$^2$ (260 mF/cm$^3$) and retains 48% capacitance at 1 V/s (Figure 8b). The areal capacitance of the nanofoam composite is found to be 3.7 mF/cm$^2$ at 0.1 V/s. While running for 10000 charge-discharge cycles, the symmetric device delivered 71% specific capacitance with respect to the first cycle (Figure 8c). The inset of Figure 8c shows the charge-discharge profile of the device found to have poor Coulombic efficiency compared to our Ti-based nanofoam composite shown above. The poor electrochemical stability and Coulombic efficiency of vanadium-based materials is one of the common problems. To evaluate charge-storage kinetics, the impedance spectra of the nanofoam composite is recorded and shown in Figure 8d. The negative phase angle of nanofoam is found to be 51° at 120 Hz. Whereas the specific capacitance of C/VO$_x$N$_y$-2 at 120 Hz is found to be 132 µF/cm$^2$, which is the highest among all studied nanofoam composites. The performance can be improved further by tuning the elemental compositions again as like C/TiO$_x$N$_y$ nanofoam composite or post-treatment.

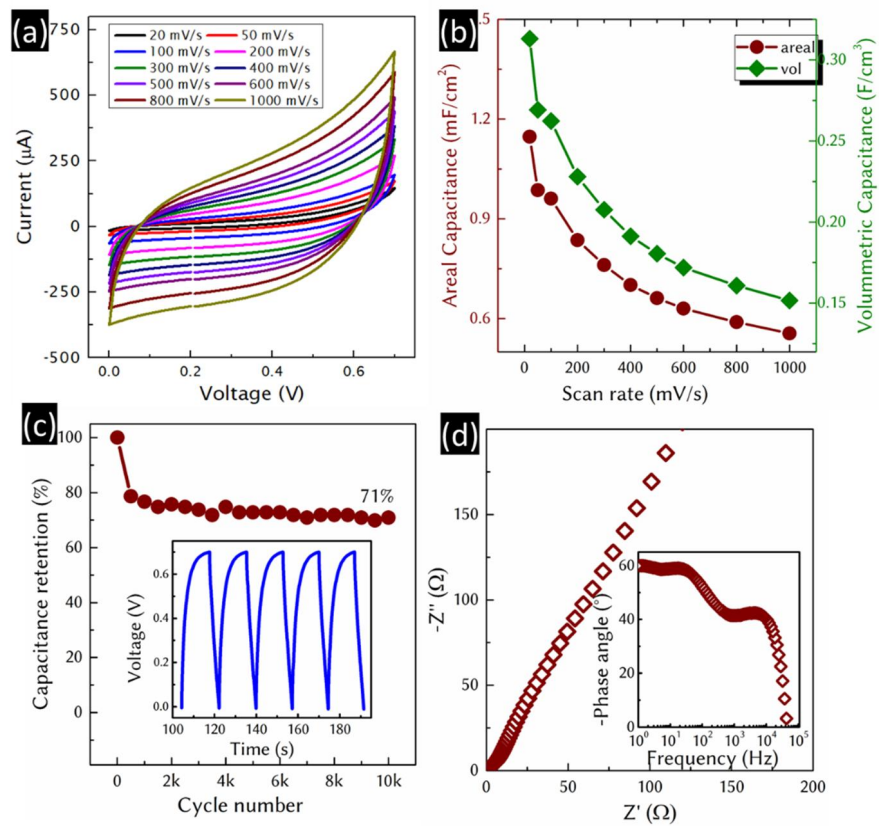

*Figure 8: Supercapacitor performance of Carbon-vanadium oxynitride.* *(a) cyclic voltammogram at different scan rate, (b) areal and volumetric capacitance with respect to the scan rate, (c) cycle stability with the charge-discharge cycles at xx current density in inset, (d) Nyquist plot with Bode plot in inset.*

## 4. Conclusion

In summary, we successfully demonstrated the co-ablation of two different targets simultaneously to synthesize the 3D porous carbon/metal oxynitride composite in one-step pulsed laser deposition with the single laser. The ratio of constituents' elements in the composites is manipulated by changing the laser ablation position and the different ratio of elements in the composites is attributed to the different sputtering yield of the target materials. Finally, we showed that the 3D porous nanocomposites have the potential to be used as binder-free and conductive agent-free electrochemical energy storage electrodes and the composites prepared by this adaptable strategy are summarized in Table 1. In comparison, the nanofoam composite with optimized constitute elements (carbon, titanium, oxygen, and nitrogen) showed the highest volumetric capacitance of 1763 mF/cm$^3$ at 0.1 V/s with high-frequency response compared to other nanofoam composites. Whereas vanadium-based nanofoam composite exhibited highest specific capacitance of 132 µF/cm$^2$ at 120 Hz. The charge-storage performance can be improved further by post-treatment of an electrode, which is the subject of our ongoing research. These results ensure that by controlling the ratio of elemental constituents of the composite, morphology, and growth rate one can optimize the structural and charge-storage properties for electrochemical energy storage applications and other desired applications.

**Authors contributions**

S.G. and C.S.C. planned and conceptualized the work. S. G., G.P. and C.H. performed the PLD deposition. S. G. and G. P. performed Raman spectroscopy of samples. S. G. M.R. and G.P. did the electrochemical measurements and assisted in the analysis. V.R. contributed to Raman spectroscopic analysis. S. G. wrote the manuscript. All authors revised the manuscript and approved the final version of the manuscript.

**Notes**

The authors declare no competing financial interest.

ACKNOWLEDGEMENT

S.G thank Horizon Europe (HORIZON) for the Marie Sklodowska-Curie Fellowship (grant no. 101067998-ENHANCER). Carlo S. Casari acknowledges partial funding from the European Research Council (ERC) under the European Union's Horizon 2020 Research and Innovation Program ERC Consolidator Grant (ERC CoG2016 EspLORE Grant Agreement 724610, website: www.esplore.polimi.it). Carlo S. Casari also acknowledges funding by the project funded under the National Recovery and Resilience Plan (NRRP), Mission 4 Component 2 Investment 1.3 Call for Tender 1561 of 11.10.2022 of Ministero dell'Università e della Ricerca (MUR), funded by the European Union NextGenerationEU Award Project Code PE0000021, Concession Decree 1561 of 11.10.2022 adopted by Ministero dell'Università e della Ricerca (MUR), CUP D43C22003090001, Project "Network 4 Energy Sustainable Transition (NEST)".

DATA AVAILABILITY: All the data of this study are available in the main manuscript and the Supplementary Information.

# SUPLLEMENTARY FILE

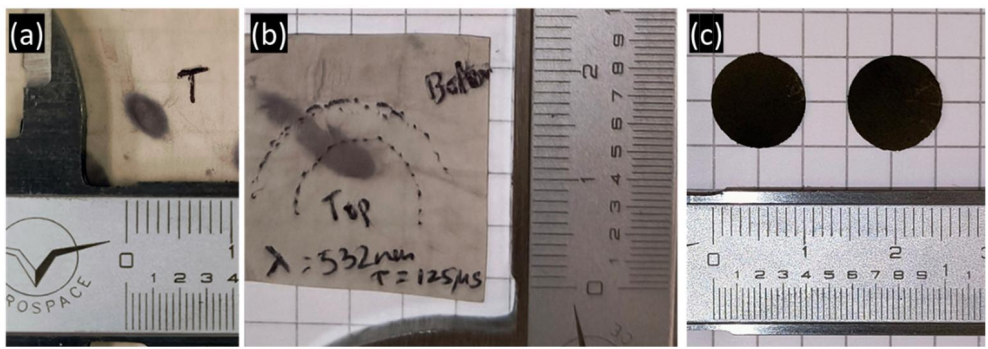

Figure S1: Laser ablation spot on the photosensitive paper placed on (a) single target material and (b) double-target material configuration shown in the schematic image of Figure 3(a) taken at ambient environment. (c) Photographic images of two symmetric electrodes of diameter of 1 cm. [Note: the ablation faded once exposed to the light. We measured the spot size first and then took the photographic image. Since photosensitive paper is exposed to light, the ablation spot is little bit faded than the spot size estimated]

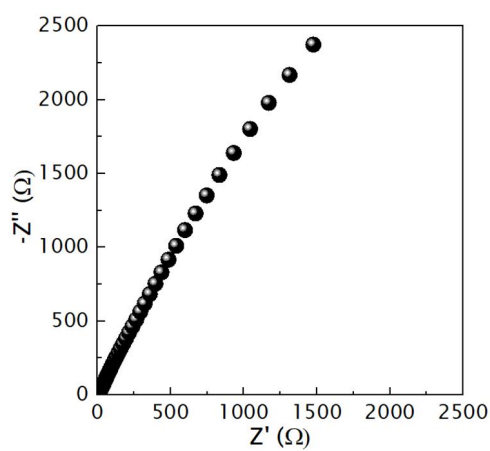

Figure S2: Electrochemical impedance spectra of C-TiOxNy

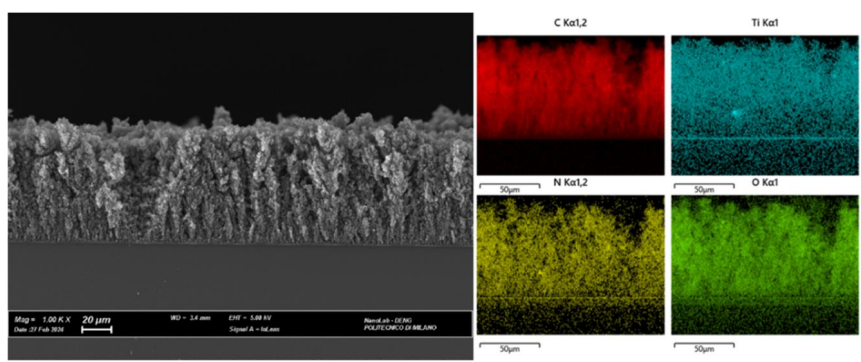

Figure S3: Cross-sectional scanning electron micrograph and elemental mapping of C-TiO$_x$N$_y$-1

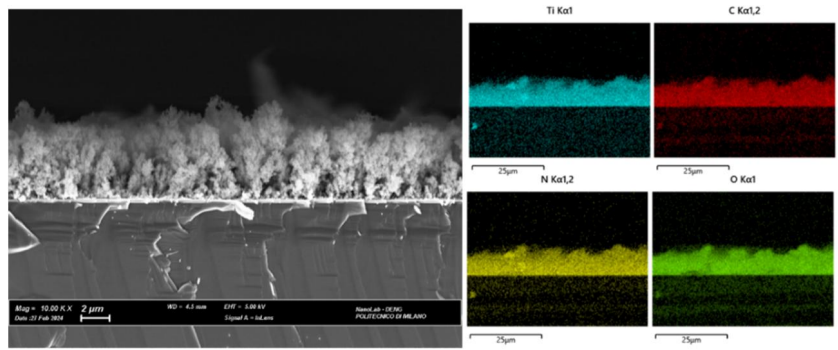

Figure S4: Cross-sectional scanning electron micrograph and elemental mapping of C-TiO$_x$N$_y$-2

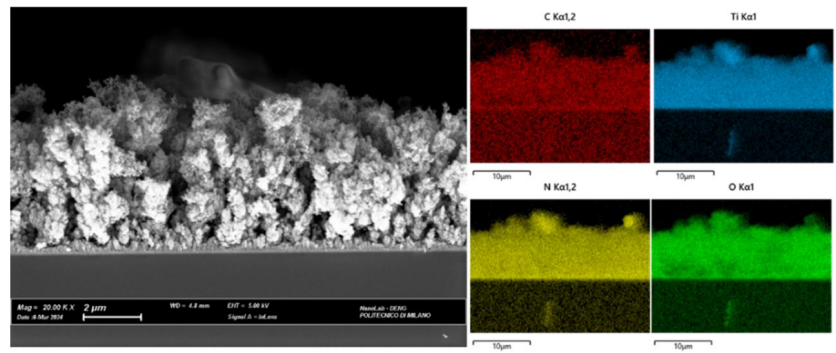

Figure S5: Cross-sectional scanning electron micrograph and elemental mapping of C/TiO$_x$N$_y$-3

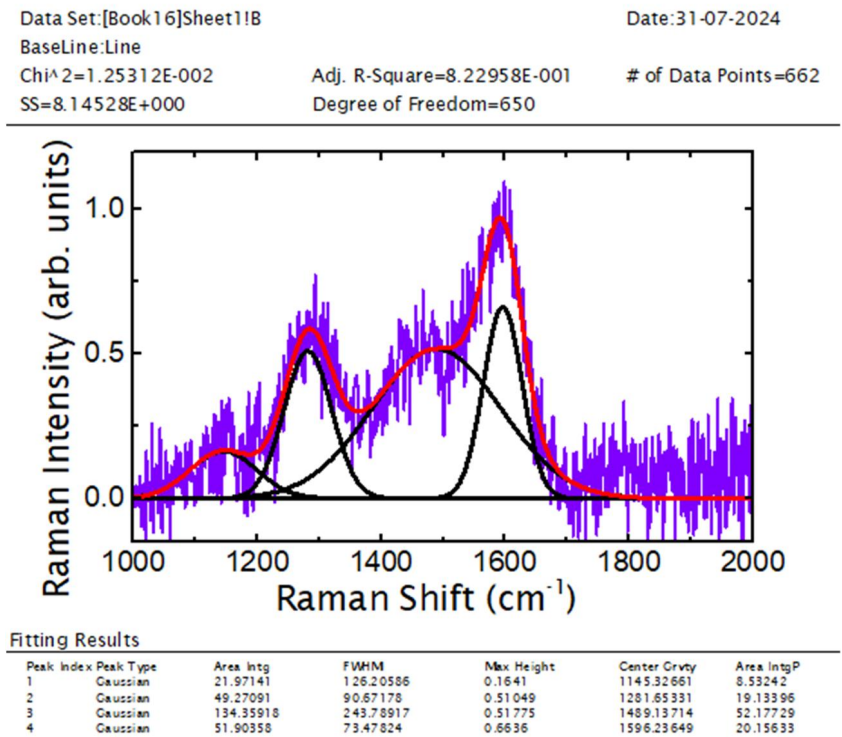

Figure S6: Deconvoluted Raman spectra of C/TiOxNy-3 in the range of 1000 to 2000 cm$^{-1}$.
21

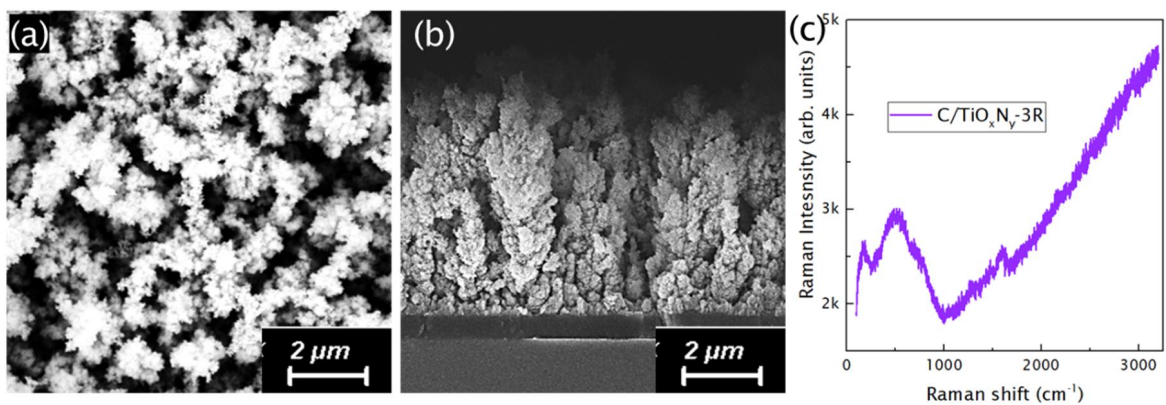

Figure S7: (a) Top-view and (b) cross-sectional view of scanning electron micrograph, and (c) Raman spectra of C/TiO$_x$N$_y$-3R.

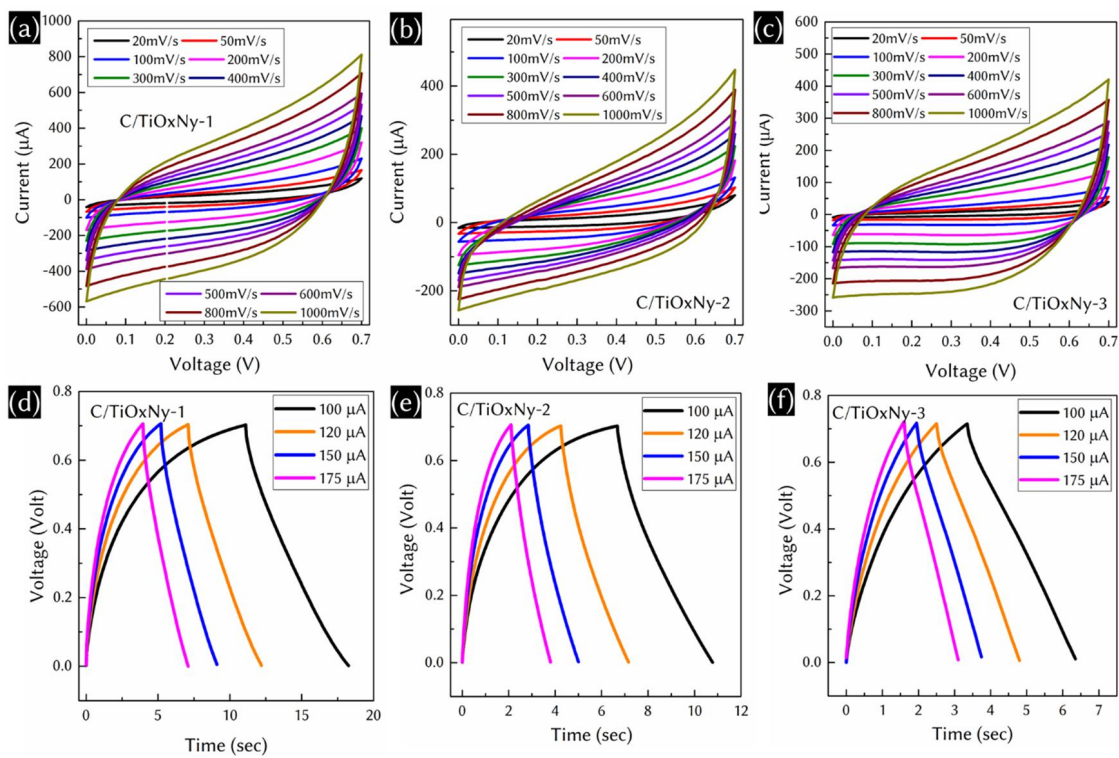

Figure S8: Cyclic voltammogram of (a) C/TiO$_x$N$_y$-1, (b) C/TiO$_x$N$_y$-2, and (c) C/TiO$_x$N$_y$-3 supercapacitor device at different scan rates. Charge-discharge profile of (a) C/TiO$_x$N$_y$-1, (b) C/TiO$_x$N$_y$-2, and (c) C/TiO$_x$N$_y$-3 supercapacitor device at different current densities.

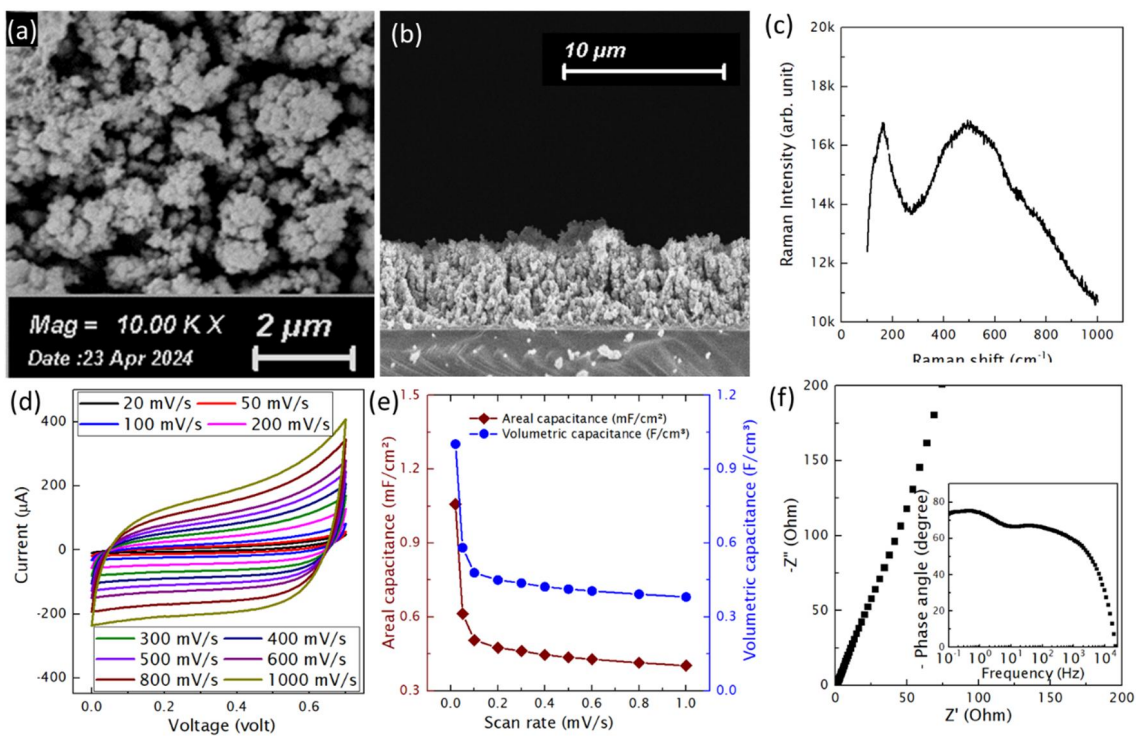

Figure S9: (a) Top view and (b) cross-sectional morphology, (c) Raman spectra of bare $TiO_xN_y$ nanofoam. (d) Cyclic voltammogram, (e) areal and volumetric capacitance with respect to the scan rate, and (f) Nyquist plot with Bode plot at inset of bare $TiO_xN_y$ nanofoam supercapacitor device.

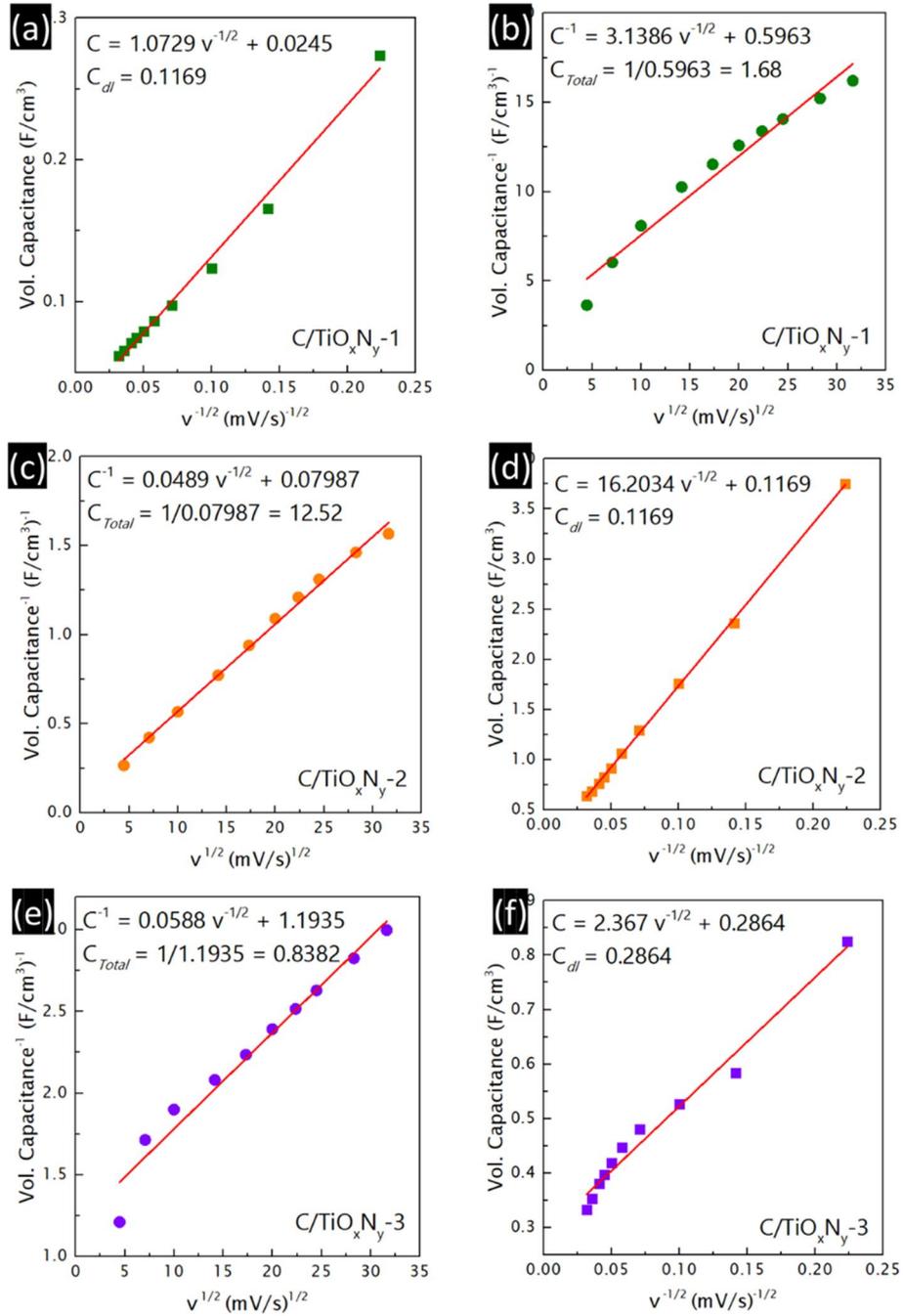

Figure S10: Plot of reciprocal volumetric capacitance versus square root of scan rate for (a) C-TiO$_x$N$_y$-1, (c) C-TiO$_x$N$_y$-2 and (e) C-TiO$_x$N$_y$-3 device. Plot of volumetric capacitance versus inverse square root of scan rate for (b), C-TiO$_x$N$_y$-1, (d) C-TiO$_x$N$_y$-2 and (f) C-TiO$_x$N$_y$-3 device.

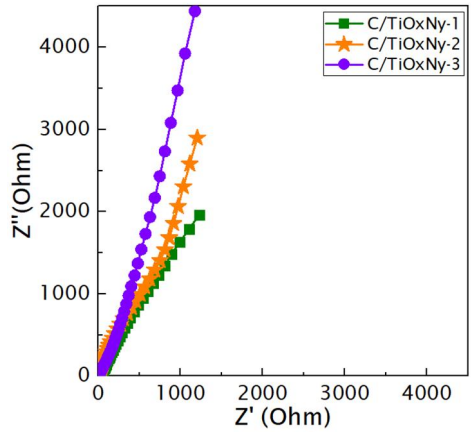

Figure S11: Electrochemical impedance spectra of C-TiO$_x$N$_y$-1 (green color), C-TiO$_x$N$_y$-2 (orange color), and C-TiO$_x$N$_y$-3 (purple color)

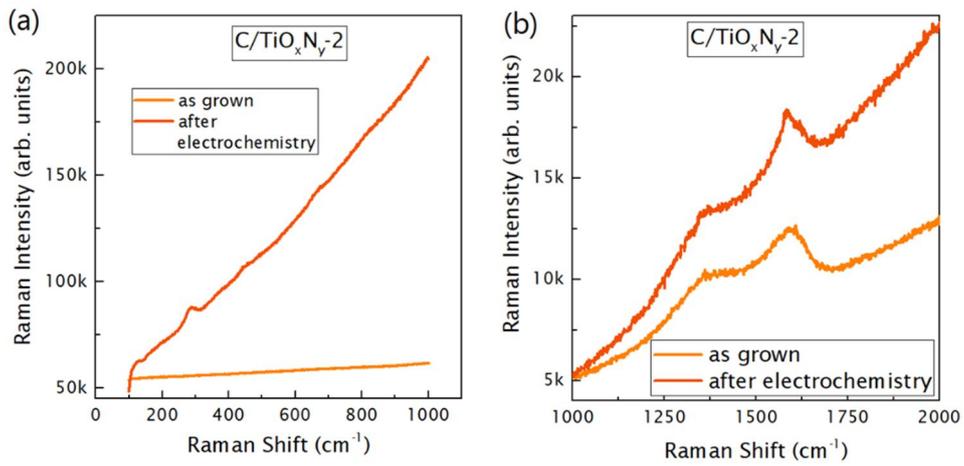

Figure S12: Raman spectra of C/TiOxNy-2 after electrochemistry with spectra of as-grown for comparison. Spectra of (a) oxynitride and (b) carbon region.

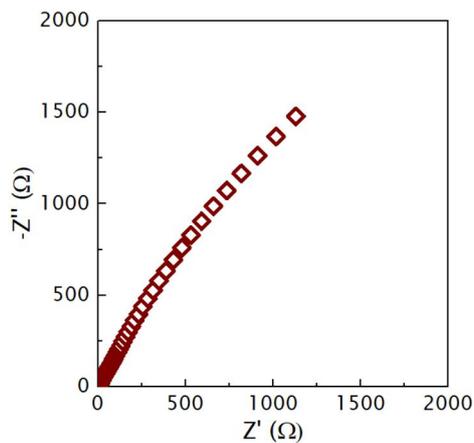

Figure S13: Electrochemical impedance spectra of C-VO$_x$N$_y$